\documentclass{edp-conf}
\usepackage{graphicx}
\begin{document}

\TitreGlobal{SF2A 2005}

%%-----------------------------
%%      the top matter
%%-----------------------------
\title{Transport and mixing by internal waves in stellar interiors: effect of the Coriolis force}
\author{Mathis, S.}\address{Observatoire de Gen\`eve; 51, chemin des maillettes, CH-1290 Versoix, Switzerland}
\secondaddress{LUTH, Observatoire de Paris, 92 195 Meudon, France}
\author{Zahn, J.-P. $^2$}
%\address{LUTH, Observatoire de Paris, 92 195 Meudon, France}
%
\runningtitle{Internal waves in stellar interiors: effect of the Coriolis force}
\setcounter{page}{1}
% Keep this line, even if the page will be settled afterwards..
\index{Mathis, S.}
\index{Zahn, J.-P.}
% Repeat the authors here, this will help to make the final index

\maketitle
\begin{abstract}
We briefly recall the physical background of the transport of angular momentum and the mixing of chemicals inside stellar radiation zones and its importance for stellar evolution. Then, we describe its present modeling, its successes and its weaknesses. Next, we introduce the new theoretical developments that allow us to treat in a self-consistent way the effect of the Coriolis force on the low-frequencies internal waves and its consequences for the transport processes. This research is aimed at improving the modeling of stellar interiors in the perspective of future astero and helioseismology missions such as COROT and GOLF-NG.
\end{abstract}
%
%%-----------------------------
%%      your text
%%-----------------------------

\section{Rotational mixing}
In standard models of stellar interiors, radiation zones which are convectively stable are postulated to be without motion other than rotation. But various observational results (i.e. surface abundances, helioseismology) show that these regions are the seat of mild mixing. The most likely cause of such mixing is the differential rotation, which drives a large scale meridional circulation due to thermal imbalance and angular momentum transport and gives probaly rise to shear instability (cf. Mathis, Palacios \& Zahn 2004). This process has been described in a self-consistent way, namely taking into account the transport of angular momentum which modifies the rotation profile (Zahn 1992, Maeder \& Zahn 1998, Mathis \& Zahn 2004). Series of models have been built which include these processes, called rotational mixing (cf. Talon et al. 1997, Meynet \& Maeder 2000), and for massive stars, they agree rather well with the observations.\\
However, rotational mixing alone cannot explain the almost uniform rotation of the radiative interior of the Sun, as revealed through helioseismology, and one must conclude therefore that another, more powerful process is operating, at least in slow rotators. The most plausible candidates are magnetic torquing (Garaud 2002) and momentum transport by internal waves (Talon et al. 2002, Talon \& Charbonnel 2005).\newpage

\par Concerning the waves, their treatment presents two major weaknesses. The first one is our crude description of their generation by turbulent convection. The second one is that our present description does not take into account the action of rotation on the waves. In this work, we have undertaken to improve the modeling of the transport by internal waves by introducing the effects of the Coriolis force. Indeed, the low-frequency internal waves which are responsible for the deposit or the extraction of angular momentum (cf. Talon et al. 2002) are strongly influenced by the rotation because the frequencies of the waves are of the same order that the inertial frequence, $2\Omega$. Thus, internal waves become gravito-inertial waves (cf. Dintrans \& Rieutord 1999) and we have to treat the action of the rotation on the waves and their feed-back on its profile.

\section{Transport of angular momentum by gravito-inertial waves}
 
The velocity field with respect to an inertial frame is $
\vec V\left(\vec r,t\right)=r\sin\theta\Omega\left(r,\theta\right){\bf \widehat e}_{\varphi}+\vec u\left(r,\theta,\varphi,t\right)$, where $\Omega\left(r,\theta\right)$ and $\vec u$ are respectively the differential rotation and the velocity of the waves. $r,\theta,\varphi$ are the classical spherical coordinates and $t$ is the time. We solve the momentum equation
\begin{equation}
D_{t}\vec u+\left[2\Omega{\bf \widehat{e}}_{z}\wedge\vec u+r\sin\theta\vec u\cdot\vec\nabla\Omega{\bf \widehat{e}}_{\varphi}\right]=-\frac{\vec\nabla P^{'}}{\rho}+\frac{\rho^{'}}{\rho}\vec g\hbox{ }\hbox{ }\hbox{ }\hbox{where}\hbox{ }D_{t}=\partial_{t}+\Omega\partial_{\varphi}, 
\end{equation}
the continuity equation
\vskip -2pc
\begin{equation}
D_{t}\rho+\vec\nabla\cdot\left(\rho\vec u\right)=0
\end{equation}
and the energy equation which is given in the adiabatic case by
\begin{equation}
D_{t}\left(\frac{\rho^{'}}{\rho}-\frac{1}{\Gamma_{1}}\frac{P^{'}}{P}\right)+\left[\frac{{\rm d}\ln \rho}{{\rm d}r}-\frac{1}{\Gamma_{1}}\frac{{\rm d}\ln P}{{\rm d}r}\right]u_{r}=0.
\end{equation}
$\vec g$ is the gravity; $\rho$ and $P$ are the density and the pressure and $\rho^{'}$ and $P^{'}$ their respective eulerian fluctuations.
We treat the case of weakly differential shellular rotation:
\begin{equation}
\Omega\left(r,\theta\right)=\overline{\Omega}\left(r\right)=\overline{\Omega}_{s}+\delta\overline{\Omega}\left(r\right)\hbox{ }\hbox{ }\hbox{where}\hbox{ }\hbox{ }\delta\overline{\Omega}\left(r\right)<\!\!\!<\overline{\Omega}_{s}.
\end{equation}
$\overline{\Omega}_{s}$ is the mean rotation rate of the region where the waves are excited by turbulent convection (namely the border between the radiation zone and the convective core or envelope which is studied), while $\delta\overline{\Omega}\left(r\right)$ is the residual differential rotation. We introduce the characteristic frequencies of the problem:
\begin{equation}
\sigma\left(r\right)=\sigma_{0}+m\overline{\Omega}\left(r\right)=\sigma_{s}+m\delta\overline{\Omega}\left(r\right)\hbox{ }\hbox{ }\hbox{where}\hbox{ }\hbox{ }\sigma_{s}=\sigma_{0}+m\overline{\Omega}_{s};
\end{equation}
$\sigma\left(r\right)$ is the local frequency ``seen" by the Doppler shifted wave, and $\sigma_{0}$ and $\sigma_{s}$ are the frequencies of the waves respectively in an inertial frame and in the corotating frame. We ignore the perturbation of the gravitational potential (Cowling approximation) and  filter out the effects of the centrifugal force, which have to be taken into account in rapid rotators.\\

We are interested here in the regime where both $2 \Omega \ll N$ and $\sigma \ll N$, and we make  the {\bf traditional approximation}, which consists in neglecting the horizontal component of the rotation vector in the momentum equation. We can then  separate the variables in the partial differential equations which govern the problem, as illustrated here for the vertical component of the velocity field:  $u_{r}=\sum_{k,m}u_{r;k,m}\left(r\right)\Theta_{k,m}\left(x;\nu\right)\exp\left[i m \varphi\right]\exp\left[i \sigma_{0}t\right]$ where $\nu=2\overline{\Omega}_{s}/\sigma_{s}$ and $x=\cos\theta$. The radial functions $u_{r;k,m}$ obey the following vertical differential equation
\begin{equation}
\frac{{\rm d}^2\Psi_{k,m}}{{\rm d}r^2}+\left[\left(\frac{N^2}{\sigma_{s}^{2}}-1\right)\frac{\Lambda_{k,m}\left(\nu\right)}{r^2}\right]\Psi_{k,m}=0\hbox{ }\hbox{where}\hbox{ }\Psi_{k,m}=\rho^{1/2}r^{2}u_{r;k,m},
\label{Vert}
\end{equation}
while the horizontal functions $\Theta_{k,m}\left(x;\nu\right)$, which are called the Hough functions, are the solutions of the Laplace equation
\begin{equation}
\left[\frac{{\rm d}}{{\rm d}x}\!\left(\!\frac{1-x^{2}}{1-\nu^{2}x^{2}}\frac{{\rm d}}{{\rm d}x}\!\right)\!-\!\frac{1}{1-\nu^2 x^2}\!\left(\!\frac{m^{2}}{1-x^{2}}\!+\!m\nu\frac{1+\nu^2x^2}{1-\nu^{2}x^{2}}\right)\!\right]\!\Theta_{k,m}=-\Lambda_{k,m}\left(\nu\right)\Theta_{k,m} ,
\end{equation}
which satisfy the relevant boundary conditions.
Those equations have the same structure that in the case where Coriolis force is ignored. The difference is the replacement of the classical horizontal eigenvalue of the spherical harmonics, $l\left(l+1\right)$, by the horizontal eigenvalue related to the Hough functions, $\Lambda_{k,m}\left(\nu\right)$. Note that the vertical wavenumber $k^{2}_{V;k,m}=\left[\left(N^2/\sigma\left(r\right)^{2}-1\right)\cdot\Lambda_{k,m}\left(\nu\right)/r^2\right]$ now depends on rotation, and therefore also the radiative damping. The second equation describes  how the horizontal functions are modified by rotation.

Next, using the continuity equation and the quasi-adiabatic approximation (cf. Press 1981) to describe the damping acting on the waves, we get the following equation for the transport of angular momentum in the hydrodynamical case:
\begin{equation}
\rho {{\rm d} \over {\rm d}t} (r^2\overline{\Omega})=\frac{1}{5r^2}\partial_{r}\left(\rho r^4\overline{\Omega}U_{2}\right)+\frac{1}{r^2}\partial_{r}\left(\rho\nu_{v}r^4\partial_{r}\overline{\Omega}\right)-\frac{1}{r^2}\partial_{r}\left[r^{2}\mathcal{F}_{J}\left(r\right)\right]\hbox{ }.
\end{equation}
The terms in the right-hand side of the equation are the divergence of respectively the mean flux of angular momentum advected by the meridional circulation, the mean viscous flux related to the shear-induced turbulence and the mean flux transported by the waves, $\mathcal{F}_{J}$, which is given by:
\begin{eqnarray}
&\!\!& 4\pi r^2 \mathcal{F}_{J}\left(r\right)\!\!=\!\!%\underbrace{
\frac{1}{2}r_{c}k_{c}^{-1}\rho_{c}{V}_{c}^{3}\frac{N_{c}}{\omega_{c}^{2}}\!\!\int_{\omega_{c}}^{N_{c}}\!\!\sum_{l=0}^{E\left[l_{u}\left(\omega\right)\right]}\!\!\sum_{m=-l}^{l}\left\{\left[\frac{l\left(l+1\right)}{\left(2l+1\right)}\right]\left(1-\frac{\omega^{2}}{N_{c}^{2}}\right)^{1\over2}\left(\frac{\omega}{\omega_{c}}\right)^{-5}\right.%}_{\hbox{Convection properties}}%
\nonumber\\
& &{\left. \cdot \sum_{\left\{k|\Lambda_{k,m}\left(\nu\right)\ge0\right\}}\left\{%\underbrace{
\left[\mathcal{P}_{l,m}^{k}\left(\nu\right)\right]^{2}\frac{\mathcal{J}_{1;k,m}\left(\nu\right)}{\Lambda_{k,m}^{1/2}\left(\nu\right)}%}_{\hbox{Coriolis force effect}}
\exp\left[-%\underbrace{
\tau_{k,m}\left(r,\delta\overline{\Omega}\left(r\right);\nu\right)%}_{\hbox{Radiative damping}}
\right]\right\}\right\}}\frac{{\rm d}\omega}{\omega}
\end{eqnarray}
where
\begin{equation}
\tau_{k,m}\left(r,\delta\overline{\Omega}\left(r\right);\nu\right)=\Lambda_{k,m}^{3\over2}\left(\nu\right)\int_{r}^{r_{c}}\left\{K\frac{N^{2}}{\sigma^{4}\left(r^{'}\right)}\left[N^{2}-\sigma^{2}\left(r^{'}\right)\right]^{1\over2}\right\}\frac{{\rm d}r^{'}}{{r^{'}}^{3}}\hbox{ } .
\end{equation}
$r_{c}$ is the radial coordinate of the top of the radiation zone,  $\rho_{c}$ and $N_{c}$ are the density and the Brunt-V\"ais\"al\"a frequency. $V_{c}$, $\omega_{c}$ and $k_{c}$ characterize the largest convective eddies which are at the origin of the generation of the waves. The $l$ are the order of spherical harmonics which describe turbulent convection on the sphere. Here, we must underline that we have taken the simplest description of turbulence, namely Kolmogorov's law (cf. Garcia Lopez \& Spruit 1991, Zahn et al. 1997). $\mathcal{P}_{l,m}^{k}\left(\nu\right)$ is the integral related to the projection of the convective velocity field on the Hough functions while $\mathcal{J}_{1;k,m}\left(\nu\right)$ is related to the mean flux of angular momentum transported by waves on an isobar. Finally, $K$ is the thermal diffusivity.

Therefore, the Coriolis force is acting on the structure of horizontal eigenfunctions of the waves but also on their radiative damping. Indeed, the values of $\Lambda_{k,m}\left(\nu\right)$ are greater for retrograde waves than for prograde ones (cf. Lee \& Saio 1997). Thus, the properties of extraction and deposit of momentum are modified. In the same way, one should also note that the values of $\mathcal{P}_{l,m}^{k}\left(\nu\right)$ are different for prograde and retrograde waves. Therefore, Coriolis force modifies the respective generation of retrograde and prograde waves by turbulent convection.

This treatment is the first one that allows a coherent treatment of the transport of angular momentum by gravito-inertial waves. We first treat the action of the Coriolis force on the waves and next their feed-back on the residual differential rotation.\\

Work is in progress to implement the new equations in existing stellar structure codes (STAREVOL \& Geneva, CESAM) to model the evolution of rotating stars and to improve the modeling of generation of internal waves by turbulent convection (spectrum and associated flux).

\end{document}